\documentclass[prb,twocolumn,superscriptaddress,showpacs,preprintnumbers,amsmath,amssymb]{revtex4}

\usepackage{graphicx}
\usepackage{dcolumn}
\usepackage{bm}

\begin{document}


\title{Correlated flux noise and decoherence in two inductively coupled flux qubits}

\author{F. Yoshihara}
\email{f-yoshihara@zb.jp.nec.com}
\affiliation{The Institute of Physical and Chemical Research (RIKEN), Wako, Saitama 351-0198, Japan}

\author{Y. Nakamura}
\affiliation{The Institute of Physical and Chemical Research (RIKEN), Wako, Saitama 351-0198, Japan}
\affiliation{NEC Nano Electronics Research Laboratories, Tsukuba, Ibaraki 305-8501, Japan}

\author{J. S. Tsai}
\affiliation{The Institute of Physical and Chemical Research (RIKEN), Wako, Saitama 351-0198, Japan}
\affiliation{NEC Nano Electronics Research Laboratories, Tsukuba, Ibaraki 305-8501, Japan}

\date{\today}

\begin{abstract}
We have studied decoherence in a system where two Josephson-junction flux qubits share a part of their superconducting loops and are inductively coupled.
By tuning the flux bias condition, we control the sensitivities of the energy levels to flux noises in each qubit.
The dephasing rate of the first excited state is enhanced or suppressed depending on the amplitudes and the signs of the sensitivities.
We have quantified the $1/f$ flux noises and their correlations and found that the dominant contribution is by local fluctuations.

\end{abstract}

\pacs{03.67.Lx,85.25.Cp,74.50.+r}
\keywords{Josephson devices, decoherence, $1/f$ noise}

\maketitle

The presence of low-frequency flux noise in superconducting devices has been known for decades.
It was discovered as an excess noise in SQUIDs \cite{Koch83IEEE,Wellstood87APL} and has recently been intensively studied as a source of dephasing in various types of superconducting qubits, such as charge, \cite{Ithier05} flux, \cite{Fum06PRL,Kakuyanagi07PRL,0812Lanting,Harris08,Harris09} and phase qubits. \cite{Claudon06,Bialczak07PRL,Bennett09}
The noise spectrum typically follows 1/$f$ frequency dependence with a spectral density of about 1-40 $\mu\Phi_0 /\mathrm{Hz}^{1/2}$ at 1 Hz, where $\Phi_0$ is the superconducting flux quantum.
The noise spectrum depends only weakly on geometry---samples with a loop area of 1 $\mathrm{\mu m}^2$ up to a few mm$^2$ are within the above range---and does not show a clear dependence on the material used.
However, the microscopic origin of the noise has been elusive so far.
It is crucial to identify and eliminate the source of such noise in order to improve the performance of these devices; i.e., the sensitivity of SQUIDs and coherence of qubits.

A few recent experimental observations have suggested as the source of the flux noise a high density of electron spins existing on the surface of superconducting electrodes.
Rogachev {\it et al.}\ attributed the magnetic-field-induced enhancement of the critical current in superconducting nanowires to suppression of spin-flip scattering at the surface, \cite{Rogachev06PRL} while Sendelbach {\it et al.}\ directly measured  the magnetization of SQUIDs at low temperatures as well as correlated inductance fluctuations. \cite{Sendelbach08PRL,Sendelbach09PRL}
Electron spin resonance studies on Si/SiO$_2$ interfaces \cite{Schenkel06} as well as scanning-SQUID measurements on Au surfaces \cite{Bluhm09} have also indicated the presence of electron spins.
As the microscopic origins of these surface spins, theoretical studies have proposed localized electrons at disordered interfaces between surface oxides and metals/semiconductors \cite{Sousa07PRB,Faoro08PRL,Choi09} or at defects in the surface oxides. \cite{Koch07PRL}

In the present study, we approached this issue by means of dephasing measurements in coupled flux qubits.
Decoherence in coupled qubits depends on the correlations between noises in each qubit. \cite{Governale01,Storcz03,You05,Hu07,DArrigo08}
It can therefore be used to characterize the noise correlations without the need for measuring the correlations directly.
In the echo decay signal of the qubits, we observed the contributions of pure dephasing due to $1/f$ flux noises and evaluated the correlations.
The results indicated local rather than global flux fluctuations in accordance with the surface spin model.

\begin{figure}
\includegraphics[width=\linewidth]{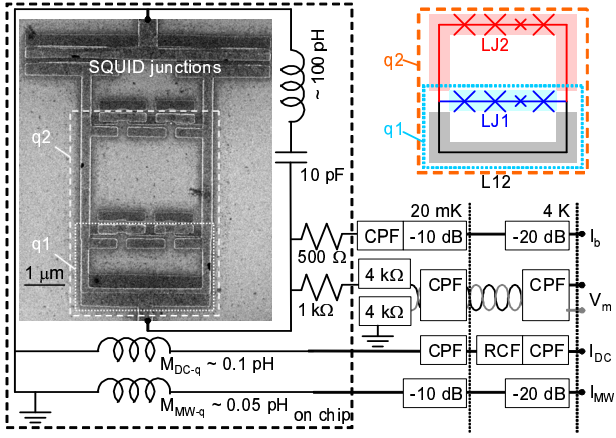}
\caption{(color online) Scanning electron micrograph of the sample and a sketch of the measurement setup. Two coupled qubits (q1 and q2) share a part of their loops with a readout SQUID shunted by an on-chip capacitor.
A current bias line ($\rm{I_b}$) and a voltage measurement line ($\rm{V_m}$) consisting of lossy shielded cables are connected to the SQUID via on-chip resistors.
Microwave current pulses are fed through an on-chip control line ($\rm{I_{MW}}$) inductively coupled to the qubit.
A local flux control line ($\rm{I_{DC}}$) is also inductively coupled to the qubit and  is connected to a battery-powered current source.
The sample chip is enclosed in a copper-shielded box, and all the wires are electrically shielded as well.
CPF and RCF denote a copper-powder filter and RC low-pass filter, respectively.
Global magnetic flux bias is applied with an external superconducting coil connected to a battery-powered current source.
The sample is cooled to 20 mK in a dilution refrigerator magnetically shielded with three $\mu$-metal layers at room temperature.
Inset: Schematic of two coupled qubits.
This system consists of three segments: a part of q1 (LJ1), a part of q2 (LJ2) and a common part (L12).
}
\label{SEM2q}
\end{figure}

The experiments were carried out using a sample fabricated by electron-beam lithography and shadow evaporation of Al films, with a 20 nm-thick first layer and 30 nm-thick second layer, on a non-doped Si wafer with a 300 nm-thick SiO$_2$ layer (Fig.$~\ref{SEM2q}$).
The qubits formed a small superconducting loop intersected by four Josephson junctions, among which one was smaller than the others by a nominal factor of 0.55.
The sample consisted of two flux qubits, q1 and q2, coupled with each other via kinetic inductance of the shared part.
The area ratio of q1 and q2 was designed to be 1:3 so that the two qubits could be closely biased to their half-integer flux-bias points simultaneously by a global magnetic field.
The conditions were $\Phi_{\rm ex1}/\Phi_0=0.5$ and $\Phi_{\rm ex2}/\Phi_0=1.5$,
where $\Phi_{\mathrm{ex}_j}$ is the externally applied flux through qubit $j$.
An additional on-chip local flux bias line made of Al allowed independent control of the flux biases in each qubit.

The effective Hamiltonian of a system with two inductively coupled flux qubits can be expressed in terms of persistent current basis as
\begin{equation}\label{Hamiltonian}
    \mathcal{H} = -\frac{1}{2}\sum_{j=1}^2(\varepsilon_j\sigma_{zj}+\Delta_j\sigma_{xj})+J_{12}\sigma_{z1}\sigma_{z2},
\end{equation}
where $\sigma_{xj}$ and $\sigma_{zj}$ are Pauli matrices,
$\Delta_{j}$ is the tunnel splitting between the two states with opposite directions of persistent current along the loop, $\varepsilon_j = 2I_{pj}\Phi_0n_{\phi j}$ is the energy bias between the two states, and $I_{pj}$ is the persistent current along the qubit loop.
The qubits are coupled with a coupling energy $J_{12}=M_{12}I_{p1}I_{p2}$, where $M_{12}$ is the mutual inductance.
We define the normalized magnetic flux in each qubit loop $n_{\phi j}$ as $\{(\Phi_{\mathrm{ex}_j} - 0.5 \Phi_0) \bmod \Phi_0\}/\Phi_0$.
In the case of an isolated qubit, $E_{01}=\Delta_j$ and $\partial E_{01}/\partial n_{\phi j}=0$ at $n_{\phi j}=0$, where $E_{01}$ is the eigenenergy of the first excited state.
This is the optimal flux bias condition where dephasing due to fluctuations of $n_{\phi j}$ is minimal.

\begin{figure}
\includegraphics[width=0.8\linewidth]{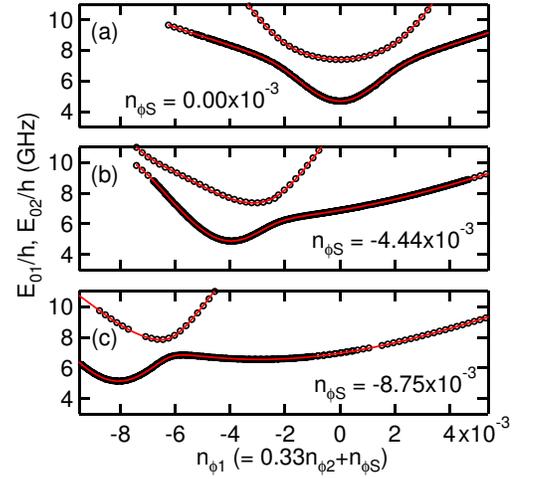}
\caption{(color online) Spectrum of the two-qubit system as a function of the global flux bias. Panels (a)-(c) are for different flux bias offsets $n_{\phi S}$.
The open circles are measured data corresponding to $E_{01}$ and $E_{02}$, the energies of the first and second excited states.
The red solid curves represent calculated energy levels using the two-qubit Hamiltonian.
}
\label{PFBs}
\end{figure}

In our experimental setup, the energies of the first and second excited states, $E_{01}$ and $E_{02}$, in the two-qubit system are measured by spectroscopy.
A 5 $\mathrm{\mu s}$ microwave pulse is applied to the system, followed by a bias current pulse of the readout SQUID.
When the microwave frequency hits a transition of the system, the excitation is detected as a change in the SQUID switching probability $P_{\rm sw}$. \cite{Chiorescu03Sci}
Figure~$\ref{PFBs}$ shows the results as a function of the global flux bias.
By sweeping the global flux bias, $n_{\phi 1}$ and $n_{\phi 2}$ vary according to the equation $n_{\phi 1} = 0.33 n_{\phi 2}+n_{\phi S}$, where $n_{\phi S}$ is an offset between the optimal flux bias points of the two qubits.
For $n_{\phi S} = 0$, the optimal bias points of q1 and q2 are aligned, while the optimal point of q2 shifts toward the left with decreasing $n_{\phi S}$.
The energy levels show anticrossings due to the strong inductive coupling between the qubits.

\begin{figure}
\includegraphics[width=0.8\linewidth]{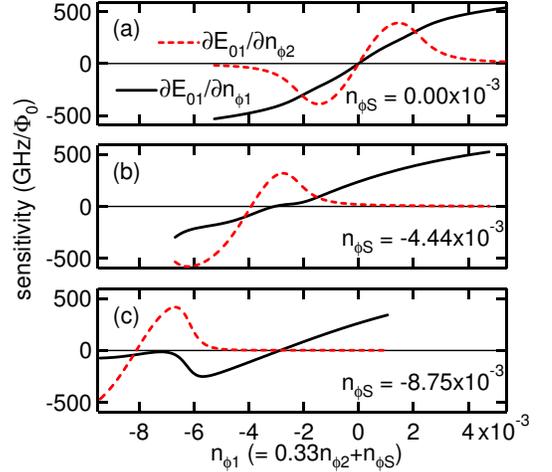}
\caption{(color online) Sensitivities of eigenenergies $E_{01}$ to flux noise in each qubit loop as a function of the global flux bias. Panels (a)-(c) are for different flux bias offsets $n_{\phi S}$.
The black solid curves are for $\partial E_{01}/\partial n_{\phi 1}$, and the red dotted curves for $\partial E_{01}/\partial n_{\phi 2}$.
}
\label{Senses}
\end{figure}

The data fit well with the calculated eigenenergies, indicating that the system can be simply described with two qubits and a fixed coupling between them.
The fitting parameters obtained from the least-squares method are as follows: $\Delta_1/h=6.56\pm0.02$ GHz, $\Delta_2/h = 5.29\pm0.03$ GHz, $I_{p1}=125\pm1$ nA, $I_{p2}=136\pm1$ nA, and $J_{12}/h=-1.20\pm0.02$ GHz.
By partially differentiating $E_{01}(n_{\phi 1},n_{\phi 2})$ with respect to $n_{\phi 1}$ and $n_{\phi 2}$, we also obtain energy sensitivities of the first excited state in the two-qubit system to flux noises in q1 and q2; i.e., $\partial E_{01}/\partial n_{\phi 1}$ and $\partial E_{01}/\partial n_{\phi 2}$ (Fig.~\ref{Senses}), respectively.

\begin{figure}
\includegraphics[width=0.8\linewidth]{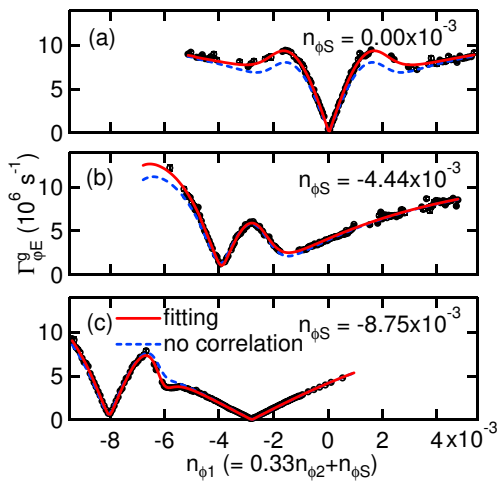}
\caption{(color online) Pure dephasing rates (circles) determined by spin-echo measurements as a function of the global flux bias. Panels (a)-(c) are for different flux bias offsets $n_{\phi S}$.
The red solid curves represent a fitting considering the amplitudes of the flux noises in each qubit and their correlations.
The blue dotted curves were calculated by subtracting the contribution of the correlations from the red solid curves.
}
\label{Gecho}
\end{figure}

Dephasing of the first excited state is characterized by Hahn echo measurement with a sequence of $(\pi/2)$-$(\pi)$-$(\pi/2)$ pulses. \cite{Hahn50PR}
The echo decay curve $P_{\rm sw}(t)$ contains both relaxation component $\exp(-\Gamma_1 t/2)$ and pure dephasing component $\langle e^{i\varphi (t)}\rangle_E$ such that $P_{\rm sw}(t) \propto \exp(-\Gamma_1 t/2)\langle e^{i\varphi (t)}\rangle_E$.
The first term is determined independently by energy relaxation measurement, while the latter is fitted well by a Gaussian decay, $\exp\{-(\Gamma_{\varphi E}^g t)^2\}$.\cite{Fum06PRL}
Figure~$\ref{Gecho}$ shows $\Gamma_{\varphi E}^g$ obtained for three different $n_{\phi S}$.
The sample has relatively good coherence: the smallest dephasing rate $\Gamma_{\varphi E}^g = 0.153 \times 10^6$~s$^{-1}$ is observed for $n_{\phi S}=-8.75\times10^{-3}$ and at $n_{\phi 1}=-2.85\times 10^{-3}$, where $\Gamma_1 = 0.148 \times 10^6$~s$^{-1}$.
The decay of the echo signal by a factor of $1/e$ takes place in  $T_{2{\rm echo}} = 4.4$~$\mu \mathrm{s}$.

We have also examined the SQUID-bias-current dependence of $\langle e^{i\varphi (t)}\rangle_E$ and confirmed that the noise from the bias current line does not significantly contribute to the pure dephasing. \cite{Fum06PRL}
Because of the strong $n_{\phi j}$ dependence of $\Gamma_{\varphi E}^g$, we can also rule out possible contributions of charge and critical-current noise to the dephasing.

The observed Gaussian decay implies the presence of low-frequency energy fluctuations with a 1/$f$ spectrum,
\begin{eqnarray}
  \nonumber
  S_{E_{01}}(\omega) &=& \frac{1}{2\pi} \int^{\infty}_{-\infty}d\tau\langle \delta E_{01}(t)\delta E_{01}(t+\tau)\rangle\exp{(-i\omega \tau)} \\
   & \equiv & \frac{A_{E_{01}}}{|\omega|}.
\end{eqnarray}
It also requires that the high-frequency tail extends up to the angular frequency range comparable to $\Gamma_{\varphi E}^g$, which is $\sim 2\pi \times 10$~MHz in the present sample.
The echo dephasing rate is then expressed as $\Gamma^g_{\varphi E}=\sqrt{A_{E_{01}}\ln2}/\hbar$.  \cite{Cottet02PHD}

As the origin of the energy fluctuations, it is most natural to assume the presence of 1/$f$ flux noises in each qubit: $S_{n_{\phi j}}(\omega) = A_{n_{\phi j}}/|\omega|$ ($j$=1,2). \cite{Fum06PRL}
Here we also consider cross correlations between the two flux fluctuations, $\delta n_{\phi 1}(t)$ and $\delta n_{\phi 2}(t)$.
Following the linear cross approximation discussed in Ref. 24,
\begin{eqnarray}
  \nonumber
  S_{n_{\phi 1}n_{\phi 2}}(\omega) &=& \frac{1}{2\pi}\int^{\infty}_{-\infty}d\tau\langle \delta n_{\phi 1}(t)\delta n_{\phi 2}(t+\tau)\rangle\exp{(-i\omega \tau)} \\
   & \equiv & \frac{A_{n_{\phi 1}n_{\phi 2}}}{|\omega|}.
\end{eqnarray}
This form of cross-correlation spectrum is justified when the $1/f$ noise is produced by many two-level systems and their coupling strengths to the qubits and switch rates are uncorrelated.
For $1/f$ charge noise, such cross correlations in charge noise have been observed using two single-electron transistors.\cite{Zorin96}
Now we fit the observed Gaussian decay rate $\Gamma_{\varphi E}^g$ using
\begin{eqnarray}
    \nonumber
    \Gamma^g_{\varphi E} &=& \frac{1}{\hbar} \left[ \ln 2 \left\{ A_{n_{\phi 1}} \left( \frac{\partial E_{01}}{\partial n_{\phi 1}} \right)^2 + A_{n_{\phi 2}} \left( \frac{\partial E_{01}}{\partial n_{\phi 2}} \right)^2 \right. \right. \\
    & & + \left. \left. 2 A_{n_{\phi 1}n_{\phi 2}} \left( \frac{\partial E_{01}}{\partial n_{\phi 1}} \right) \left( \frac{\partial E_{01}}{\partial n_{\phi 2}}\right) \right\} \right]^{1/2}
\end{eqnarray}
(red solid curves in Fig.~\ref{Gecho}).
All the data points of $\Gamma_{\varphi E}^g$ for different flux biases are fitted with a single set of parameters,
$A_{n_{\phi 1}}=[(2.32 \pm 0.01) \times 10^{-6}]^2$, $A_{n_{\phi 2}}=[(2.76 \pm 0.01) \times 10^{-6}]^2$, and $A_{n_{\phi 1}n_{\phi 2}}=[(1.49 \pm 0.01) \times 10^{-6}]^2$, which confirms the validity of our assumption of dephasing induced by $1/f$ flux noises.
Moreover, the noise amplitudes, similar to the previously reported values, are determined with high accuracy.
We also note that the $\Gamma_{\varphi E}^g$ data cannot be fitted well if $A_{n_{\phi 1}n_{\phi 2}}$ is set to zero.

The small ratios $A_{n_{\phi 2}}/A_{n_{\phi 1}}=1.42$ and $A_{n_{\phi 1}n_{\phi 2}}/A_{n_{\phi 1}}=0.42$ are clear proofs that global flux noise is not a dominant source for the dephasing in the first excited state: global flux noise would give rise to $A_{n_{\phi 2}}/A_{n_{\phi 1}}=9$ and $A_{n_{\phi 1}n_{\phi 2}}/A_{n_{\phi 1}}=3$ due to the difference between the areas of q1 and q2.

On the other hand, we obtain a non-negligible amount of $A_{n_{\phi 1}n_{\phi 2}}$, which indicates noticeable correlations between $\delta n_{\phi 1}(t)$ and $\delta n_{\phi 2}(t)$.
In Fig.~\ref{Gecho}, dephasing rates calculated without taking into account the correlation term are plotted as blue dotted curves.
The observed dephasing rate is enhanced (suppressed) when sensitivities $\partial E_{01}/\partial n_{\phi 1}$ and $\partial E_{01}/\partial n_{\phi 2}$ have the same (opposite) signs.
For example, at around $n_{\phi 1}=-6\times 10^{-3}$ for $n_{\phi S}=-8.75\times 10^{-3}$, reduction of the dephasing rate due to the correlation is observed.

To qualitatively account for the flux noise amplitudes $A_{n_{\phi 1}}$, $A_{n_{\phi 2}}$, and especially the correlation term $A_{n_{\phi 1}n_{\phi 2}}$, we introduce a simple model where flux noises are generated by a number of  microscopic sources, described as fluctuating magnetic dipoles, scattered over the sample.
The two-qubit system can be divided into three segments (see the inset of Fig.$~\ref{SEM2q}$): a part of the q1 loop and its junctions (LJ1), a part of the q2 loop and its junctions (LJ2), and the common part of the two loops (L12).
Coupling between each dipole and a qubit reaches a maximum when the dipole sits on the surface of the superconducting loops, while dipoles away from the surface do not couple effectively. \cite{Koch07PRL}
For such sources on the surface, we further apply the following approximation:
Sources along LJ1 (LJ2) couple exclusively to q1 (q2) and those along L12 couple equally to both q1 and q2.
For the latter, $\delta n_{\phi 1}$ and $\delta n_{\phi 2}$ have full correlations.

Summing up all the contributions of dephasing from LJ1, LJ2, and L12, $A_{E_{01}}$ can be rewritten as
\begin{eqnarray}
\nonumber  A_{E_{01}} & = & A_{n_{\phi}}^{\mathrm{LJ1}} \left( \frac{\partial E_{01}}{\partial n_{\phi 1}} \right)^2 + A_{n_{\phi}}^{\mathrm{LJ2}} \left( \frac{\partial E_{01}}{\partial n_{\phi 2}} \right)^2 \\
  & & +A_{n_{\phi}}^{\mathrm{L12}} \left( \frac{\partial E_{01}}{\partial n_{\phi 1}} + \frac{\partial E_{01}}{\partial n_{\phi 2}} \right)^2.
\label{eq:AE01}
\end{eqnarray}
The noise amplitudes in each part of the system are calculated as $A_{n_{\phi}}^{\mathrm{LJ1}}=(1.77 \times 10^{-6})^2$, $A_{n_{\phi}}^{\mathrm{LJ2}}=(2.32 \times 10^{-6})^2$, and $A_{n_{\phi}}^{\mathrm{L12}}=(1.49 \times 10^{-6})^2$, which represent the distribution of flux noise sources on the loops.
Two points are worth mentioning here:
(i) There are two parts of the system, (L12 + LJ1) and LJ2, each consisting of four Josephson junctions and a similar length of superconducting loops.
The noise amplitudes in the (L12 + LJ1) part are calculated as $A_{n_{\phi}}^{\mathrm{LJ1}} + A_{n_{\phi}}^{\mathrm{L12}} = (2.32 \times 10^{-6})^2$, which coincides with $A_{n_{\phi}}^{\mathrm{LJ2}}$.
This result is consistent with the model and the assumption of the local flux noises.
(ii) The third term in Eq.(\ref{eq:AE01}) originating from L12 can be cancelled when the sensitivities satisfy the equation, $\partial E_{01}/\partial n_{\phi 1} = -\partial E_{01}/\partial n_{\phi 2}$.

In conclusion, we have studied dephasing in two inductively coupled flux qubits.
The dominant source of the dephasing is found to be low-frequency $1/f$ flux noises.
At the same time, the local, rather than global, nature of the flux noises is revealed in the correlations between the noises in each qubit.
The results agree with a model in which flux noise sources are distributed on the surface of superconducting loops.

We are grateful to L. Ioffe, M. R\"{o}tteler, J. Pekola, W. Oliver, E. Paladino, and P. Billangeon for their valuable discussions, and to K. Harrabi for assistance with the experiments.
This work was supported by the CREST program of the Japan Science and Technology Agency (JST) and the Grant-in-Aid for Scientific Research Program for Quantum Cybernetics of the Ministry of Education, Culture, Sports, Science and Technology (MEXT), Japan.


\end{document}